\documentclass[prl,twocolumn,superscriptaddress,showpacs,nofootinbib]{revtex4-2}
\usepackage{graphicx}
\usepackage{float}
\usepackage{dcolumn}
\usepackage{float}
\usepackage{bm}
\usepackage{mathrsfs}
\usepackage{fullpage}
\usepackage{ulem}
\usepackage[per-mode = fraction]{siunitx}
\usepackage{hyperref}
\hypersetup{
    colorlinks=true,
    linkcolor=blue,
    filecolor=magenta,      
    urlcolor=cyan,
    citecolor=magenta,
}
\usepackage{xcolor}
\usepackage{soul}
\usepackage{amsthm,amssymb}
\usepackage{mathtools}
\usepackage{comment}
\usepackage{physics}

\begin{document}


\title{A Kapitza Pendulum Route to Supercurrent Tunnel Diodes}


\author{Yuriy Yerin}
\affiliation{Istituto di Struttura della Materia of the National Research Council, via Salaria Km 29.3, I-00016 Monterotondo Stazione, Italy}
\affiliation{\mbox{Institute for Theoretical Solid State Physics, IFW Dresden, 01069 Dresden, Germany}}
\author{Stefan-Ludwig Drechsler}
\affiliation{\mbox{Institute for Theoretical Solid State Physics, IFW Dresden, 01069 Dresden, Germany}}
\author{A. A. Varlamov}
\affiliation{CNR-SPIN, via del Fosso del Cavaliere, 100, 00133 Roma, Italy}
\affiliation{Istituto Lombardo ``Accademia di Scienze e Lettere'', via Borgonuovo,
25 - 20121 Milan, Italy}
\author{Francesco Giazotto}
\affiliation{NEST Istituto Nanoscienze-CNR and Scuola Normale Superiore, I-56127, Pisa, Italy}
\author{Jeroen van den Brink}
\affiliation{\mbox{Institute for Theoretical Solid State Physics, IFW Dresden, 01069 Dresden, Germany}}
\affiliation{\mbox{W\"urzburg-Dresden Cluster of Excellence ct.qmat, Germany}}
\author{Mario Cuoco}
\affiliation{CNR-SPIN, c/o Universit\'a di Salerno, I-84084 Fisciano (SA), Italy}

\date{\today}

\begin{abstract}
Superconducting diodes that support nonreciprocal supercurrent flow 
{in principle constitute attractive, non-dissipative, circuit elements for superconducting electronics. But their realization faces} 
fundamental challenges, as conventional Josephson tunnel junctions are inherently reciprocal. Existing approaches to 
{break reciprocity}
typically 
{involve}
magnetism or spin–orbit coupling, which often increase device complexity and limit reproducibility. Here, we 
{demonstrate}
an alternative dynamical route to supercurrent nonreciprocity based on parametric driving. By applying a frequency-modulated supercurrent amplitude
we show that effective higher-order, nonharmonic terms are generated in the current–phase relation. 
Leveraging mathematical equivalences with the Kapitza pendulum, 
\footnote{The Kapitza pendulum is an inverted pendulum stabilized by rapid vertical oscillations of its pivot, whose slow dynamics is described by an effective time-averaged potential. The stabilization effect was first noted by
Stephenson in 1908~\cite{Stephenson}, while Kapitza's 1951 work established the standard averaging-based description~\cite{Kapitza_1951, Kapitsa_1951_ufn}.}
we show that these terms dynamically break reciprocity.
{This establishes}
the concept of a Kapitza supercurrent diode and demonstrates that nonreciprocal superconducting transport can be engineered by nonequilibrium driving
{conventional Josephson tunnel junctions.}
We propose two implementations of the Kapitza supercurrent diode—via gate-controlled superconducting interferometers or flux-driven double-loop SQUIDs—to achieve nonreciprocal supercurrent transport within experimentally accessible frequencies $\omega/2\pi \sim 1$--$10\,\mathrm{GHz}$.
\end{abstract}

\maketitle

The search for superconducting diodes, devices that allow supercurrent to flow predominantly in one direction, is an active focal point of both fundamental and technological research \cite{and20,baur22, bau22, wu22,jeo22,nad23,Ghosh2024}. 
Nonreciprocal supercurrents, or supercurrent diode effects, have been observed across a wide range of superconducting systems, including artificial superlattice \cite{and20,narita2022}, elemental \cite{Margineda2023}, noncentrosymmetric \cite{wakatsuki2017}, and polar superconductors \cite{itahashi2020}, as well as twisted graphene \cite{lin2022,Rothstein2026} and high-$T_c$ junctions \cite{G2024,Qi2025,Guo2025,Volkov_2024}, highlighting diverse strategies to break time-reversal symmetry and control directional transport \cite{and20,wakatsuki2017,zhang2020,itahashi2020,narita2022,lin2022, yerin2024,Yerin_SQUID}.
The Josephson diode effect (JDE) has been proposed and experimentally observed on different Josephson platforms, including junctions with spin–orbit-coupled weak links \cite{baumgartner2022,Costa2023,lotfizadeh2024,Borgongino2025}, Dayem bridges \cite{souto2022,MertBorzkurt2023,margineda2023sign}, and interferometric devices with higher-harmonic contributions \cite{greco2023,greco2024} and inhomogeneities \cite{Chirolli2025}. 
The JDE has further been explored in systems featuring screening currents \cite{hou2023,sundaresh2023}, self-field effects \cite{krasnov1997,golod2022}, and Josephson--current--induced phase reconfiguration mediated by kinetic inductance \cite{chen2024}. It has also been studied in multiterminal configurations \cite{gupta2023,zhang2024}, gate-tunable devices \cite{ciaccia2023,gupta2023}, as well as in vortex-driven junctions \cite{Golod2010,Golod2015,golod2022,GutfreundNatComms2023,Ji21,Fukaya2025} and through back-action mechanisms \cite{de2024quasi}.

Despite these numerous proposals for the different types and mechanisms behind superconducting diodes, a major challenge stems from the intrinsic characteristics of Josephson junctions.
At the heart of this challenge lies the reciprocal nature of the conventional Josephson effect, where the supercurrent depends sinusoidally on the phase difference $\varphi$ between the superconductors forming the junction, following the relation $I = I_c \sin(\varphi)$.
The parity symmetry of this relation with respect to the phase bias ensures that, without additional modifications, the supercurrent amplitude is identical in both directions, precluding any rectification or nonreciprocal behavior.
\\
Achieving nonreciprocal supercurrents in Josephson junctions requires introducing mechanisms that induce asymmetry in the current-phase relation, such as an anomalous phase shift or higher-order harmonic components. Interestingly, even when time-reversal and inversion symmetries are broken, resulting in an anomalous phase shift, denoted $\phi$, that alters the current-phase relation to $I = I_c \sin(\varphi + \phi)$, the supercurrent amplitude remains reciprocal. These modifications typically involve breaking time-reversal symmetry, for example, through magnetic fields or magnetic elements assisted by spin-orbit coupling, and typically require complex interface engineering.
The inclusion of a second harmonic term in the current phase relation, leading to $I = I_{1} \sin(\varphi + \phi)+I_{2} \sin(2 \varphi)$, is a minimal necessary ingredient to achieve nonreciprocal supercurrent flow.
However, controlling and reliably reproducing these harmonic components in conventional junctions is challenging. In particular, beyond managing the sources of time-reversal symmetry, the high-harmonic current-phase components required for diode functionality are sensitive to junction quality, interface properties, and material-specific effects, making consistent implementation intricate.

This raises a fundamental question: Can one realize nonreciprocal supercurrents using Josephson tunnel junctions with a standard current-phase relation? 
Addressing this question is crucial for advancing superconducting diode technology and understanding the fundamental physics that enable nonreciprocal superconducting transport.

To this end, we introduce a general method for inducing supercurrent nonreciprocity in Josephson tunnel junctions via parametric driving \cite{eichler2023classical}. This approach draws inspiration from the Kapitza pendulum \cite{Kapitza_1951, Kapitsa_1951_ufn}—a classical pendulum with a rapidly oscillating pivot point. As high-frequency oscillations can stabilize an inverted pendulum, our work demonstrates that parametric driving of coupled Josephson junctions can generate non-harmonic components and non-reciprocal supercurrent behavior. 
We propose a minimal setup for realizing a supercurrent diode based on superconducting interferometers composed exclusively of Josephson tunnel junctions, in which the supercurrent is parametrically driven by either electric or magnetic fields, establishing viable routes to realizing a Kapitza supercurrent diode.

We recall the fundamental principle of the Kapitza pendulum \cite{Kapitza_1951, Kapitsa_1951_ufn}, a rigid pendulum with a vertically oscillating pivot. If the suspension vibrates with frequency $\omega$ and amplitude $a$, the angle $\varphi$ obeys $\ddot{\varphi}=-(\omega_0^2+\alpha\omega^2\cos\omega t)\sin\varphi$, where $\omega_0^2=g/l$ and $\alpha=a/l$. In the regime of small amplitude ($\alpha\ll1$) and high frequency ($\omega\gg\omega_0$), the motion separates into a slow component $\xi_0$ and a fast oscillation $\xi$, with $\varphi=\xi_0+\xi$. A perturbative expansion and time averaging yield an effective equation $\ddot{\xi_0}=-\partial{\xi_0}V_{\mathrm{eff}}$, with $V_{\mathrm{eff}}=\omega_0^2\cos\xi_0+\tfrac{1}{2}\alpha\omega^2\cos2\xi_0$. The periodic drive thus generates a second-harmonic contribution to the potential \cite{Kapitza_1951, Kapitsa_1951_ufn}.
\\
It is well-known that the dynamics of a Josephson junction can be effectively described by analogy with the motion of a classical pendulum. Specifically, in the resistively ($R$) and capacitively ($C$) shunted junction (RCSJ) model, the phase difference $\varphi$ across the junction obeys the nonlinear differential equation:
$C \frac{d^2 \varphi}{dt^2} + \frac{1}{R} \frac{d\varphi}{dt} + I_c \sin(\varphi) = I_b$ 
which mirrors the equation of motion for a damped pendulum subject to an external torque $I_{\text{ext}}$.
Now, suppose that the total current-phase relation of the system consists of two contributions $I=I_1+I_2$.
The first term, $I_1=I_\text{c,1} \sin(\varphi +\phi)$, incorporates a phase shift $\phi$ arising from a source of time-reversal symmetry breaking. The second term, $I_2=I_\text{c,2} \cos(\omega t) \sin(\varphi)$, indicates a periodically modulated critical current at frequency $\omega$. 
Taking into account the analogy with the classical Kapitza pendulum, one expects that the resulting supercurrent has a nonreciprocal behavior when the amplitude of one of the Josephson channels is parametrically driven in time.

\begin{figure}
\includegraphics[width=1.1\linewidth]{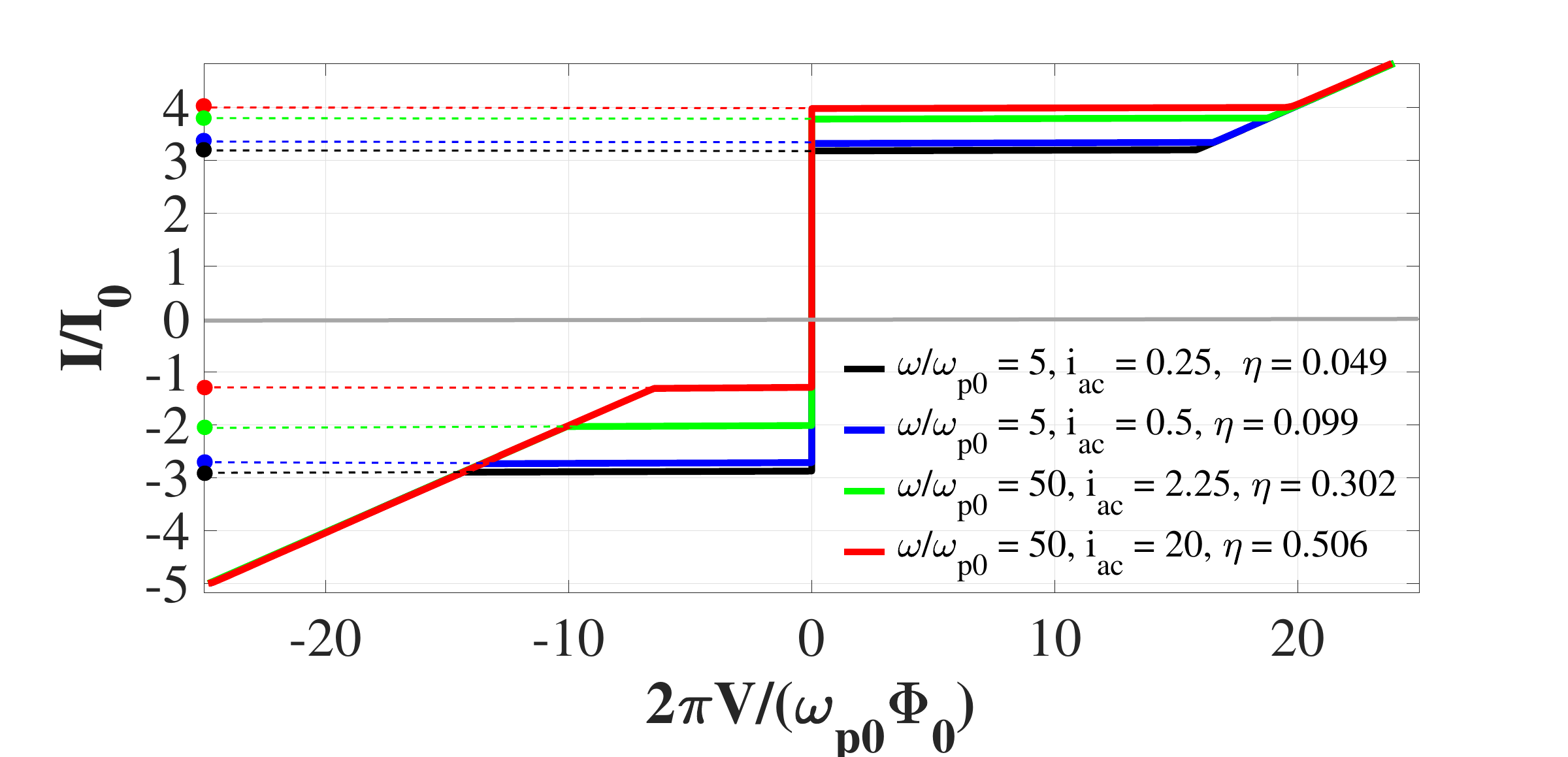}\hfil
\caption{
Current–voltage characteristics of a parametrically driven Josephson junction with anomalous phase $\phi=\pi/2$ at $i_\text{dc}=0$ for $i_\text{ac}=0.25$, $0.5$ ($\omega/\omega_{p0}=5$, black/blue), and $i_\text{ac}=2.25$, $20$ ($\omega/\omega_{p0}=50$, green/red). Inset: diode efficiency $\eta$ extracted from each curve. Filled markers indicate $I_c^{\pm}$, and color-matched dotted lines guide the eye. Damping $\beta/\omega_{p0}=0.1$, $i_{c,t}=2$.
}
\label{CVC}
\end{figure}

The equation of motion for the phase dynamics of the driven Josephson junction admits a direct interpretation as the dynamics of damped pendulum subject to a time-dependent (parametrically modulated) restoring torque. Identifying the Josephson phase $x$ with the pendulum angle $\varphi$, the equation can be expressed in a dimensionless form as
\begin{equation}
\ddot x + \beta \dot x
+ \big[i_{\mathrm{dc}}+i_{\mathrm{ac}}\cos(\omega t)\big]\sin x
+ i_{c,t} \sin(x+\phi)
= \frac{1}{2} i_b ,
\label{eq:EOM_main}
\end{equation}
where currents are normalized to a reference scale $I_0$ chosen as the characteristic Josephson critical current of the junction and time is measured in units of plasma frequency $\omega_{p0}=\sqrt{2\pi I_0/(C\Phi_0)}$ and dimensionless damping parameter $\beta = 1/(RC\,\omega_{p0})$. Eq. ~\eqref{eq:EOM_main} can be read as the equation of motion of a damped
rotor (pendulum) with angular coordinate $x(t)$ under an external torque (see the Supplemental Material~\cite{SM} for the detailed reduction of the Josephson dynamics to Eq.~\eqref{eq:EOM_main}).
The damping term $\beta\dot x$ is the usual viscous friction, while the bias $i_b$ plays the role of a constant tilting torque that favors rotation in one direction. The term $\big[i_{\rm dc}+i_{\rm ac}\cos(\omega t)\big]\sin x$ is a restoring torque whose strength is modulated over time, providing the parametric driving force. 
The additional channel $i_{c,t}\sin(x+\phi)$ is a second restoring torque that, however, is shifted in phase by $\phi$. Mechanically, this corresponds to
adding a second periodic potential whose minimum is displaced relative to the first one. 

More broadly, Eq.~\eqref{eq:EOM_main} represents a generic class of parametrically driven nonlinear phase oscillators, suggesting that the same Kapitza-type rectification mechanism may extend well beyond superconducting tunnel junctions. In particular, related implementations may be envisioned in bosonic Josephson
junctions, driven polariton weak links, spin-torque oscillators, and other nonlinear rotators with phase-shifted restoring channels.

To illustrate the emergence of nonreciprocal supercurrent behavior, we start performing numerical simulations of the driven Josephson junction (Eq. \ref{eq:EOM_main}). Here, we concentrate on a single time-dependent harmonic component ($i_{\rm ac}$) in the time-reversal symmetric contribution to the supercurrent amplitude, while the time-reversal broken contribution $i_{c,t}$ is kept constant over time for given values of the anomalous phase $\phi$. Modifying the time-dependent profile of the driving term does not alter the qualitative features of the resulting nonreciprocal behavior.
The profile of the current-voltage characteristic is reported in Fig. \ref{CVC} assuming a representative case for the anomalous phase, $\phi=\pi/2$. We find that the forward ($I^+_c$) and backward ($I^-_c$) critical currents of the zero-voltage state differ in magnitude, leading to a non-vanishing rectification amplitude defined as $\eta=\frac{I^+_c-|I^-_c|}{I^+_c+|I^-_c|}$. The rectification amplitude is found to depend on both the frequency and the amplitude of the driving mode, reaching 50$\%$ supercurrent rectification when the driving frequency significantly exceeds the plasma frequency $\omega_{p0}$. 

The frequency dependence of $I_c^+$ and $I_c^-$ indicates that, in the slow-driving regime, the two quantities are nearly identical and that there exists a threshold at $\omega/\omega_{p0} \sim 2$, above which the rectification increases significantly [Fig.~\ref{eta_crit_currents_omega}]. This trend is qualitatively reproduced when the anomalous phase is varied $\phi$. Indeed, a sizable rectification amplitude of the order of $15\%$ is obtained close to the resonance at $\omega/\omega_{p0}\sim 1$, accompanied by a decrease in amplitude when moving to higher frequency [Fig.~\ref{eta_crit_currents_omega} (c),(d)]. 


\begin{figure}
\includegraphics[width=1.1\linewidth]{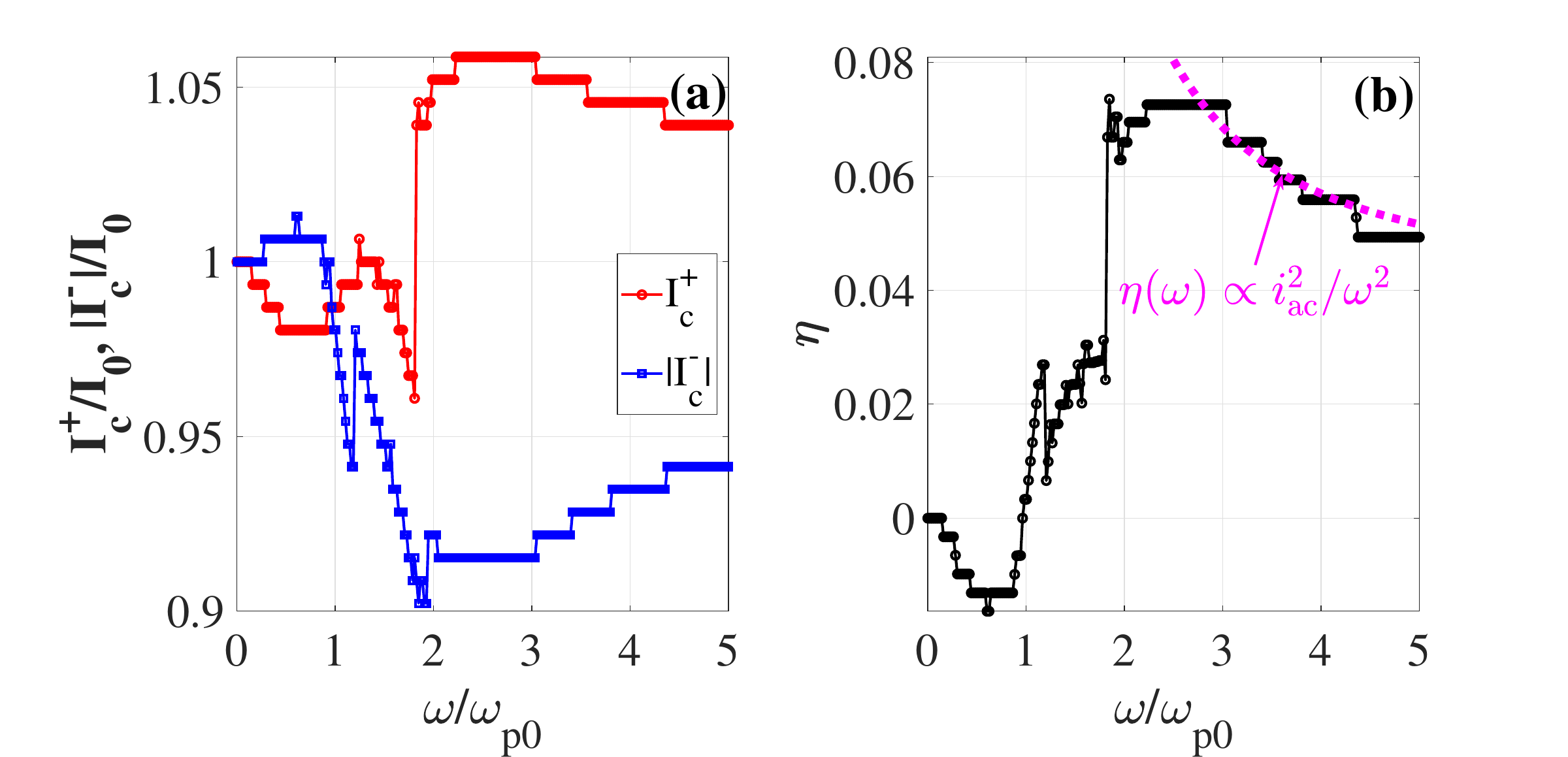}
\includegraphics[width=1.1\linewidth]{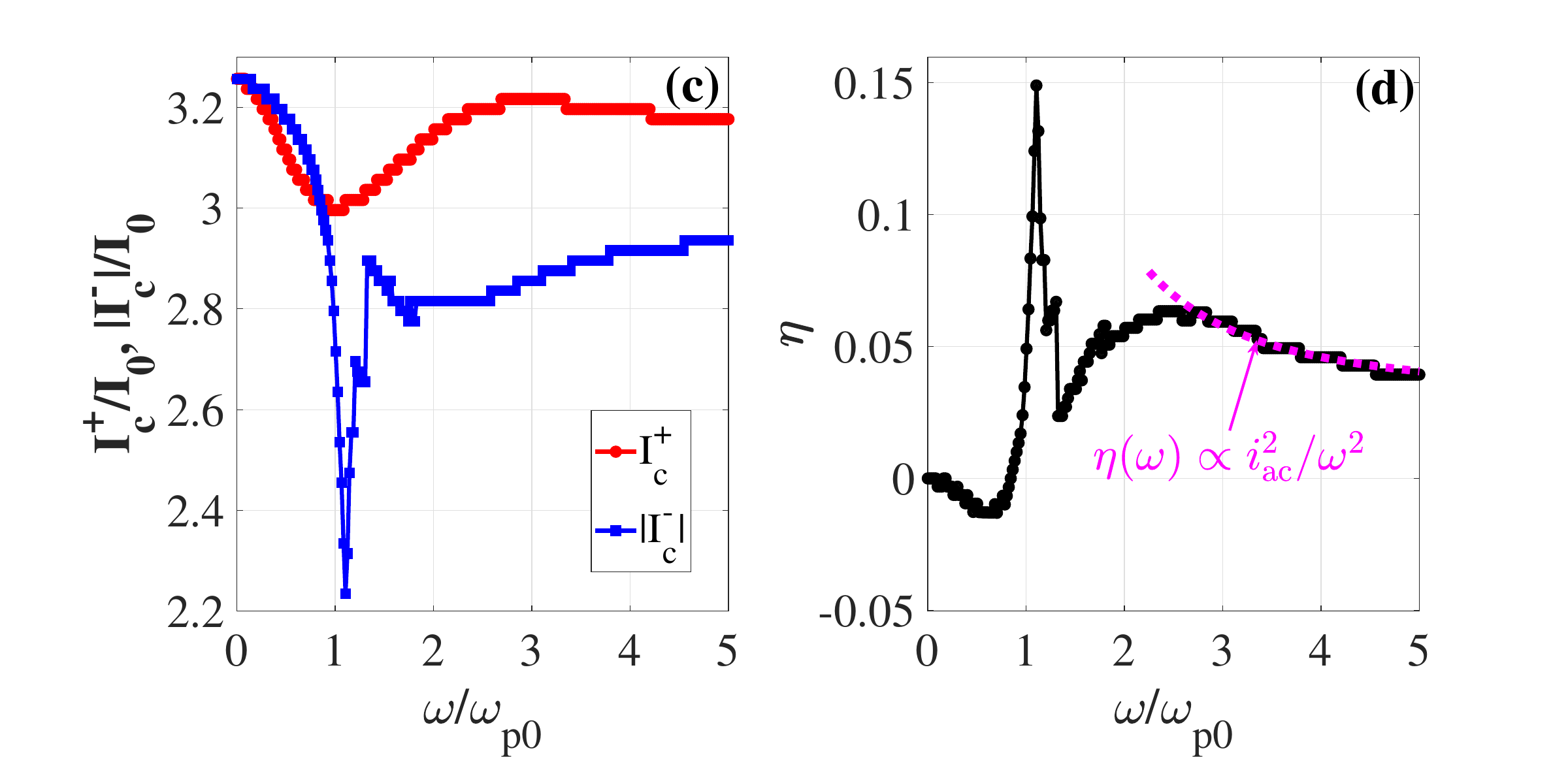}
\caption{
(a, c) Critical currents $I_c^{\pm}$ and (b, d) diode rectification $\eta$ of a driven Josephson junction versus $\omega/\omega_{p0}$ for $\phi=\pi/2$ (a, b) and $\phi=\pi/3$ (c, d) with $i_\text{dc}=0$, $i_\text{ac}=0.25$. Magenta dotted lines show high-frequency asymptotics from the weak-amplitude expansion (see main text).
}
\label{eta_crit_currents_omega}
\end{figure}

It is both instructive and useful to analyze the emergence of nonreciprocal behavior by examining the analytical expression for the frequency-dependent rectification amplitude of the driven Josephson junction within the Kapitza framework.
One can proceed by decomposing the phase as 
$x(t) = X(t) + \xi(t)$,
where $X(t)$ is a slow variable and $\xi(t)$ is a small, rapidly oscillating
correction induced by the driving field. Substituting into Eq.~\eqref{eq:EOM_main}
and expanding to the second order in $\xi$ yields
\begin{eqnarray*}
&&\ddot X + \ddot\xi
+ \beta(\dot X + \dot\xi)+
\\&&+ \big[i_{\mathrm{dc}}+i_{\mathrm{ac}}\cos(\omega t)\big]
    \big(\sin X + \xi\cos X - \tfrac{1}{2}\xi^2\sin X \big)
\nonumber\\
&&+ i_{c,t}\big(\sin(X+\phi) + \xi\cos(X+\phi) +\\ &&- \tfrac{1}{2}\xi^2\sin(X+\phi) \big)= \frac{1}{2} i_b .
\label{eq:expanded}
\end{eqnarray*}
In the spirit of the Kapitza method, the equation can be separated for the fast variable $\xi(t)$, neglecting the slow velocity and acceleration terms, $\ddot X,\dot X$, on the fast time scale. 
Then, by keeping only the terms linear in $\xi$ and proportional to the
oscillatory part of the pre-factor, $\delta c(t)\equiv i_{\mathrm{ac}}\cos(\omega t)$, we obtain
\begin{equation}
\ddot\xi + \beta \dot\xi + \omega_0^2\,\xi
\simeq -\,i_{\mathrm{ac}}\cos(\omega t)\,\sin X ,
\label{eq:xi-eq}
\end{equation}
where the local curvature (stiffness) of the effective potential is
\begin{equation}
\omega_0^2(X)
\equiv
i_{\mathrm{dc}}\cos X + i_{c,t} \cos(X+\phi).
\label{eq:omega0-def}
\end{equation}
On the fast time scale, $X$ can be considered to be approximately constant, so $\omega_0^2(X)$ is treated as a parameter. Eq. \eqref{eq:xi-eq} describes a damped harmonic oscillator driven at frequency $\omega$. The steady periodic solution can be written as
\begin{equation}
\xi(t)=\Re\!\left[\tilde\xi\,e^{i\omega t}\right],
\ \ \ 
\tilde\xi=
-\frac{i_{\mathrm{ac}}\sin X}{\omega_0^2(X)-\omega^2+i\beta\omega}.
\label{eq:xi-sol-corr}
\end{equation}

We now return to Eq.~\eqref{eq:expanded} and perform the time average over the
fast oscillations, keeping terms up to order $\xi^2$. The averaged equation for the slow variable thus becomes
\begin{eqnarray*}
&&\ddot X + \beta \dot X
+ i_{\mathrm{dc}}\sin X + i_{c,t}\sin(X+\phi) \\
&&+\ i_{\rm ac}\cos X\,\big\langle \cos(\omega t)\,\xi(t)\big\rangle_t \\
&& - \frac{1}{2}\big\langle\xi^2\big\rangle_t\,
\Big[i_{\mathrm{dc}}\sin X + i_{c,t} \sin(X+\phi)\Big]
= \frac{1}{2} i_b \,.
\label{eq:slow-eq}
\end{eqnarray*}
The term $\langle\xi^2\rangle_t$ induces an effective Josephson coupling, producing a static contribution to the current–phase relation of the form
\begin{equation}
I_s^{\mathrm{eff}}(X)
= A_1 \sin X + B_1\cos X + A_2(\omega)\,\sin 2X,
\label{eq:CPR-eff-general}
\end{equation}
with the first-harmonic coefficients
$A_1 = i_{\mathrm{dc}} + i_{c,t}\cos\phi$, and 
$B_1 = i_{c,t}\sin\phi$.
The second harmonic coefficient, instead, can be expressed as 
\begin{equation}
A_2(\omega)=
-\frac{i_{\mathrm{ac}}^2}{4}\,
\frac{\omega_0^2(x_0)-\omega^2}{\big(\omega_0^2(x_0)-\omega^2\big)^2+(\beta\omega)^2},
\label{eq:A2-corr}
\end{equation}
where $X$ has been evaluated at the static equilibrium $X=x_0$ with $x_0=-\arctan\!\frac{i_{c,t}\sin\phi}{i_{\mathrm{dc}}+i_{c,t}\cos\phi}$.
We neglect higher-order terms in $i_{\mathrm{ac}}$ and nonlinear corrections.
Introducing the first-harmonic amplitude and phase shift
$Q = \sqrt{A_1^2+B_1^2},\qquad
\delta = \arctan\!\frac{B_1}{A_1},$
we can rewrite Eq.~\eqref{eq:CPR-eff-general} as
$I_s^{\mathrm{eff}}(X)
= Q\sin(X+\delta) + A_2(\omega)\,\sin 2X $. 
For $|A_2|\ll Q$, the extrema of
$I_s^{\mathrm{eff}}(X)=Q\sin(X+\delta)+A_2\sin2X$ give
$I_c^+ \approx Q + A_2(\omega)\,\sin(2\delta),
$
$I_c^- \approx -Q + A_2(\omega)\,\sin(2\delta),$
and therefore the diode efficiency
\begin{equation}
\eta(\omega)
\;\approx\;
-\frac{i_{\mathrm{ac}}^2}{4Q}\,
\frac{(\omega_0^2(x_0)-\omega^2) \sin 2\delta}
     {\big[\omega_0^2(x_0)-\omega^2\big]^2+(\beta\omega)^2},
\,
\label{eq:eta-final-correct_main}
\end{equation}
with $\omega_0(x_0) \sim \omega_{p0}$.
From Eq. (\ref{eq:eta-final-correct_main}), one observes that, in agreement with the numerical analysis, it attains a maximum when $\omega \sim \omega_0(x_0)$. Although not vanishing, the rectification is suppressed at high frequencies and decays as $\eta(\omega)\propto i_{\mathrm{ac}}^2/\omega^2$ for $|\omega|\to\infty$ within the weak-amplitude expansion of the driving field, as one can see in Figs. \ref{eta_crit_currents_omega}b and d (magenta dotted lines). Moreover, the rectification amplitude scales with the strength of the driving field and changes sign in the vicinity of the resonance at $\omega \sim \omega_0(x_0)$. \textcolor{black}{It is important to recall that Eq. ~\eqref{eq:eta-final-correct_main} is valid in the regime ($|\xi|\ll1$), i.e., to the leading order in $i_{\rm ac}^2$; a strong-driving field with large $i_{\rm ac}$ can lead to sizable quantitative deviations due to the occurrence of higher harmonics beyond $\sin 2X$.}


\begin{figure}[h]
\includegraphics[width=0.8\linewidth]{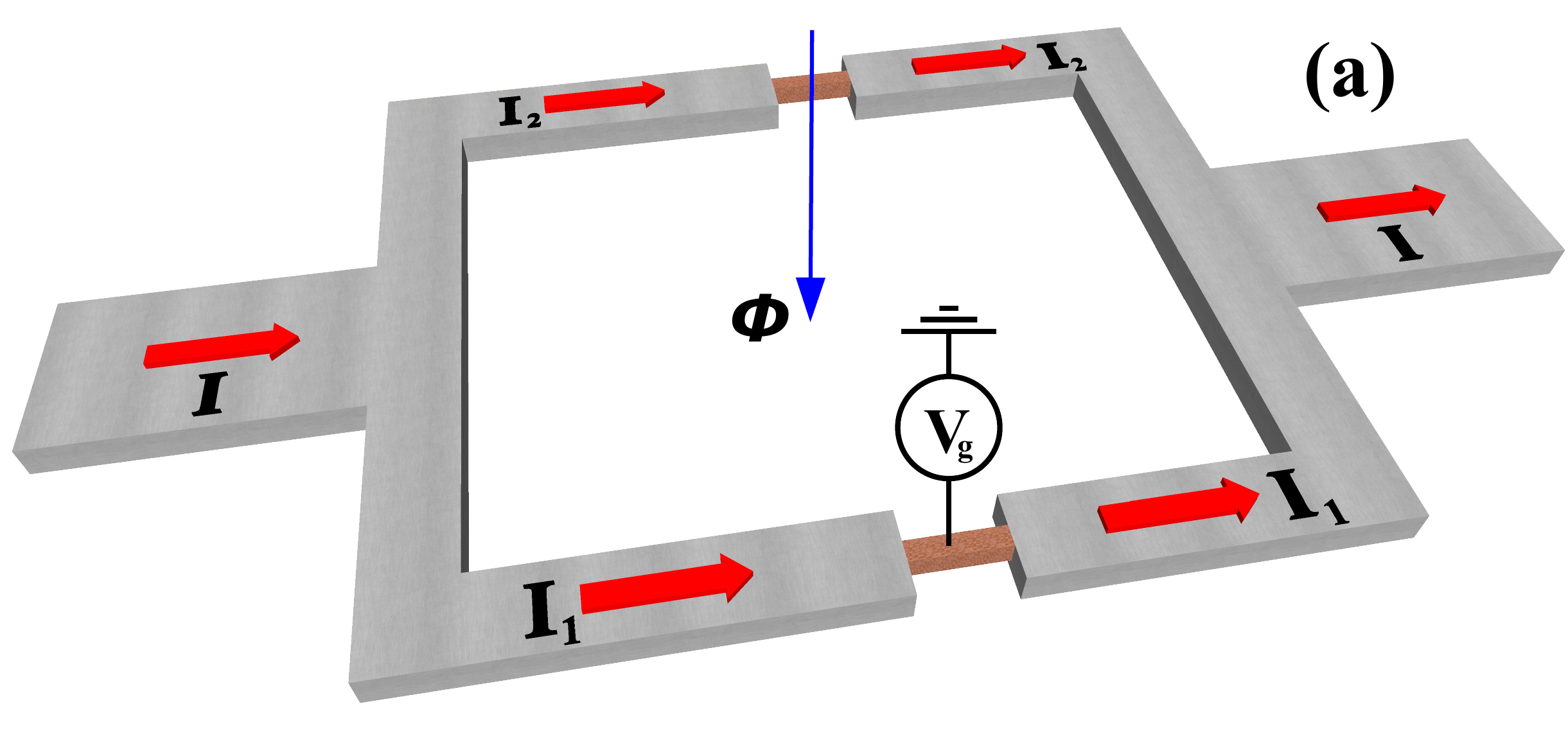}
\includegraphics[width=0.8\linewidth]{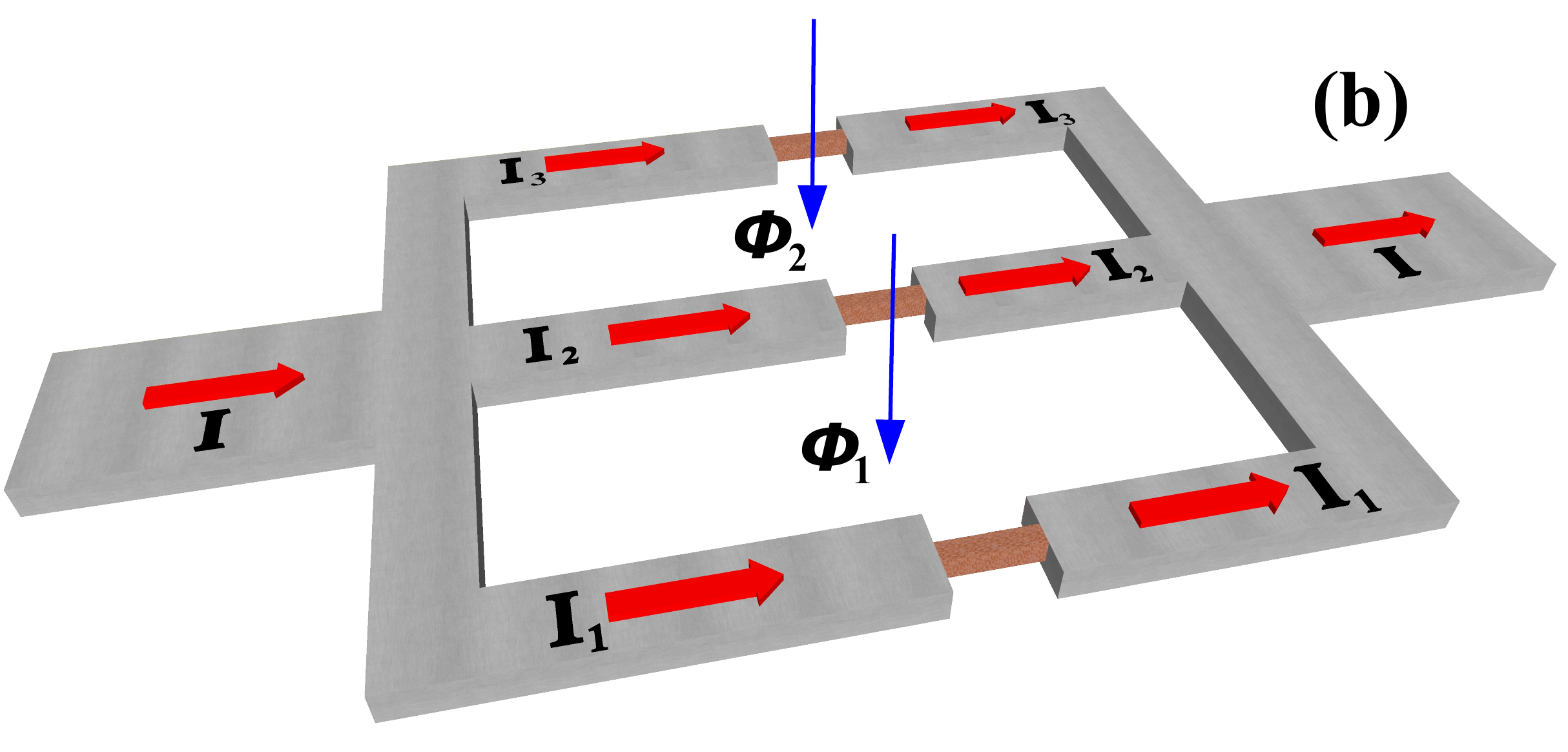}
\caption{
Schematic implementations of the Kapitza supercurrent diode using superconducting interferometers. (a) Gate-controlled SQUID, with time-dependent gate modulation (V$_g$) and a static flux ($\Phi$) -induced phase shift. (b) Double-loop interferometer with parametric flux drive and static flux breaking time-reversal symmetry. In both cases, parametric driving combined with symmetry breaking yields nonreciprocal supercurrent transport.
}
\label{model}
\end{figure}

Having established the nonequilibrium supercurrent diode effect both numerically and analytically, we now discuss possible implementations of the Kapitza supercurrent diode. Two distinct design strategies can be pursued, both relying on interferometric schemes to break the time-reversal symmetry. The first approach is based on gate-controlled supercurrents \cite{Ruf2024,DeSimoni2019,Mercaldo2020,Paolucci2019}, where the supercurrent is driven by a time-dependent gate voltage. The second relies entirely on magnetic control using superconducting interferometers, in which the parametric modulation of the supercurrent is induced by a periodically time-dependent magnetic flux.
In the first implementation [Fig.~\ref{model}(a)], time-reversal symmetry is broken using an electrically controlled superconducting interferometer in a SQUID-like geometry. The device consists of a superconducting loop interrupted by Josephson tunnel junctions whose critical currents can be tuned via electrostatic gates. A time-dependent gate voltage is applied to one of the junctions, inducing a periodic modulation of the supercurrent, while an externally applied magnetic flux is used to introduce an anomalous phase shift in the other Josephson junction \cite{SM}.

The second implementation exploits time-dependent magnetic control in a superconducting interferometer with double-loop geometry. One loop is threaded by a periodically time-dependent magnetic flux, providing the parametric drive, while the other loop is biased by a static flux and acts as a source of time-reversal symmetry breaking. 
By expressing current-phase relations in terms of average and relative phases, the dynamics of the system can be mapped to that of a driven Josephson junction with an effective static time-reversal symmetry-breaking term in the supercurrent, leading to nonreciprocal transport \cite{SM}.

The relevant frequency window for the time-periodic modulation is set by the Josephson plasma frequency,
which defines the intrinsic dynamical scale of the junction. For typical parameters ($I_c\sim 1$ nA--$5\,\mu\mathrm{A}$ and $C\sim 1$--$100\,\mathrm{fF}$) \cite{Golubov2004}, one obtains $\omega_{p0}/2\pi \sim 0.2$--$100\,\mathrm{GHz}$. To realize the proposed nonreciprocal dynamical regime, the drive frequency $\omega$ should satisfy $\omega_J \ll \Omega \lesssim 5 \omega_{p0}$, where $\omega_J$ is the characteristic Josephson frequency associated with slow phase dynamics (typically in the MHz to sub-GHz range). This condition defines an experimentally accessible window $ \omega/2\pi \sim 1$--$10\,\mathrm{GHz}$, well within the range achievable by using either microwave gate modulation or on-chip flux control in superconducting interferometers. 

Y. Y. acknowledges the funding received from HPC National Center for HPC, Big Data and Quantum Computing - HPC (Centro Nazionale 01 – CN0000013). M.C. and F.G. acknowledge support from the PNRR MUR project PE0000023-NQSTI.
M.C. acknowledges support by the Italian Ministry of University and Research (MUR) PRIN 2022 under Grant No. 2022LP5K7 (BEAT).
JvdB thanks the W\"urzburg-Dresden Cluster of Excellence on Complexity and Topology in Quantum Matter ct.qmat (EXC 2147, project-id 390858490) for support.

\bibliography{Biblio}

\clearpage
\widetext
\onecolumngrid

\renewcommand{\thefigure}{S\arabic{figure}}
\setcounter{figure}{0}
\renewcommand{\theequation}{S\arabic{equation}}
\setcounter{equation}{0}

\begin{center}
{\Large \bf Supplemental Material for}

{\Large \bf \textit{"A Kapitza Pendulum Route to Supercurrent Tunnel Diodes"}}
\end{center}

\maketitle

This Supplemental Material presents the detailed reduction of the single- and double-loop SQUID models to the effective parametrically driven pendulum equation, together with the derivation of the Kapitza-averaged current-phase relation and diode efficiency.

\section{Reduction of a gated single-loop dc SQUID at zero applied flux to an Effective Parametrically Driven Pendulum}
\label{sec:supp_singleloop_gate}

In this section we derive the effective phase equation of motion
\begin{equation}
\ddot x + \beta \dot x
+ \big[i_{\mathrm{dc}}+i_{\mathrm{ac}}\cos(\omega t)\big]\sin x
+ i_{c,t} \sin(x+\phi)
= \frac{1}{2} i_b ,
\label{eq:EOM_supp}
\end{equation}
starting from the standard RCSJ description of a single-loop dc SQUID
(two Josephson elements in parallel forming a superconducting loop as in Figure 3a of the main text), in the regime of zero applied flux and negligible loop inductance.

\subsection{Circuit model}
We consider a dc SQUID with two Josephson branches (labeled $j=1,2$), biased by a total current $I_b$. The gauge-invariant phase drops across the two branches are denoted by $\varphi_1(t_{\rm phys})$ and $\varphi_2(t_{\rm phys})$, and the voltages satisfy the Josephson relations
\begin{equation}
V_j(t_{\rm phys})=\frac{\Phi_0}{2\pi}\,\dot\varphi_j(t_{\rm phys}),
\qquad j=1,2,
\label{eq:JJ_voltage}
\end{equation}
where dot here means $d/dt_{\rm phys}$ with non-dimesionless time $t_{\rm phys}$ and $\Phi_0$ is the flux quantum.

Each branch is modeled by an RCSJ element with capacitance $C_j$ and shunt
resistance $R_j$, so that the branch current $I_j$ is
\begin{equation}
I_j
= C_j\frac{\Phi_0}{2\pi}\,\ddot\varphi_j
+ \frac{1}{R_j}\frac{\Phi_0}{2\pi}\,\dot\varphi_j
+ I_{s,j}(\varphi_j,t_{\rm phys}).
\label{eq:RCSJ_each}
\end{equation}
The total bias current splits between the two branches:
\begin{equation}
I_b = I_1+I_2.
\label{eq:Ib_split}
\end{equation}

To obtain Eq.~\eqref{eq:EOM_supp} one must assume that the total supercurrent contains two distinct sinusoidal channels: leg 1 is gate-modulated and has an ordinary CPR,
\begin{equation}
I_{s,1}(\varphi_1,t_{\rm phys})
= I_{c,1}(t_{\rm phys})\,\sin\varphi_1,
\qquad
I_{c,1}(t_{\rm phys})
= I_{c,1}^{(0)} + \delta I_{c,1}\cos(\Omega t_{\rm phys}).
\label{eq:gate_mod_Ic}
\end{equation}
while leg 2 provides a static phase-shifted CPR,
\begin{equation}
I_{s,2}(\varphi_2)
= I_{c,t}\,\sin(\varphi_2+\phi),
\label{eq:phi_shift_CPR}
\end{equation}
with constant amplitude $I_{c,t}$ and a fixed offset $\phi$. This offset can represent, e.g., an anomalous ($\varphi_0$) junction or any
built-in phase bias mechanism producing $\sin(\varphi+\phi)$ even at $\Phi=0$.

Without such a genuinely phase-shifted second channel, the second term would collapse into a renormalization of the coefficient of $\sin x$ and Eq.~\eqref{eq:EOM_supp} would not result as written.

\subsection{Zero-flux constraint and single-phase reduction}
Fluxoid quantization in a dc SQUID gives the constraint
\begin{equation}
\varphi_1-\varphi_2
= \frac{2\pi}{\Phi_0}\Big(\Phi_{\rm ext}+L I_{\rm circ}\Big),
\label{eq:fluxoid}
\end{equation}
where $\Phi_{\rm ext}$ is the applied flux, $L$ is the loop inductance, and
$I_{\rm circ}$ is the circulating current.
In the regime used in the main text,
\begin{equation}
\Phi_{\rm ext}=0,
\qquad
L\to 0,
\label{eq:zero_flux_small_L}
\end{equation}
Eq.~\eqref{eq:fluxoid} enforces $\varphi_1=\varphi_2$.
We therefore introduce a single collective phase variable
\begin{equation}
x(t_{\rm phys}) \equiv \varphi_1(t_{\rm phys})=\varphi_2(t_{\rm phys}),
\label{eq:single_phase_def}
\end{equation}
which also implies a common voltage across both branches by
Eq.~\eqref{eq:JJ_voltage}.

\subsection{Physical equation of motion for the collective phase}
Summing Eq.~\eqref{eq:RCSJ_each} over $j=1,2$, using
Eqs.~\eqref{eq:Ib_split}--\eqref{eq:single_phase_def}, we obtain
\begin{equation}
C_\Sigma\frac{\Phi_0}{2\pi}\,\ddot x
+\frac{1}{R_\Sigma}\frac{\Phi_0}{2\pi}\,\dot x
+\Big[I_{c,1}^{(0)}+\delta I_{c,1}\cos(\Omega t_{\rm phys})\Big]\sin x
+ I_{c,t}\sin(x+\phi)
= I_b,
\label{eq:phys_EOM}
\end{equation}
where the parallel combinations are
\begin{equation}
C_\Sigma \equiv C_1+C_2,
\qquad
\frac{1}{R_\Sigma}\equiv \frac{1}{R_1}+\frac{1}{R_2}.
\label{eq:Reff_Ceff}
\end{equation}

\subsection{Dimensionless units and mapping to the effective phase equation of motion}
We now nondimensionalize Eq.~\eqref{eq:phys_EOM}.
Choose a reference current scale $I_0$ (in practice one may take
$I_0=I_{c,1}^{(0)}$, or any convenient constant current used throughout the
paper). Define the plasma frequency associated with $(I_0,C_\Sigma)$,
\begin{equation}
\omega_{p0}\equiv
\sqrt{\frac{2\pi I_0}{\Phi_0\,C_\Sigma}}.
\label{eq:wp0_def}
\end{equation}
Introduce the dimensionless time $t=\omega_{p0}t_{\rm phys}$, so that
$d/dt_{\rm phys}=\omega_{p0}\,d/dt$.
Define dimensionless parameters
\begin{equation}
\beta \equiv \frac{1}{R_\Sigma C_\Sigma \omega_{p0}},
\qquad
\omega\equiv \frac{\Omega}{\omega_{p0}},
\label{eq:beta_omega_def}
\end{equation}
and dimensionless currents
\begin{equation}
i_{\rm dc}\equiv \frac{I_{c,1}^{(0)}}{I_0},
\qquad
i_{\rm ac}\equiv \frac{\delta I_{c,1}}{I_0},
\qquad
i_{c,t}\equiv \frac{I_{c,t}}{I_0}.
\label{eq:dimensionless_Ic}
\end{equation}
Finally, to match the convention used in Eq.~\eqref{eq:EOM_supp} (and in the
numerical code, where the bias enters as $+\tfrac12 i_b$), we define
\begin{equation}
i_b \equiv \frac{2 I_b}{I_0}
\qquad \Longleftrightarrow \qquad
\frac{I_b}{I_0}=\frac12\,i_b.
\label{eq:ib_convention}
\end{equation}
Dividing Eq.~\eqref{eq:phys_EOM} by $I_0$ and rewriting in the dimensionless
time $t$ yields exactly
\begin{equation}
\ddot x + \beta \dot x
+ \big[i_{\mathrm{dc}}+i_{\mathrm{ac}}\cos(\omega t)\big]\sin x
+ i_{c,t} \sin(x+\phi)
= \frac{1}{2} i_b ,
\label{eq:EOM_final_supp}
\end{equation}
which is Eq.~\eqref{eq:EOM} of the main text.

\subsection{Mechanical interpretation}
Equation~\eqref{eq:EOM_final_supp} can be viewed as a damped pendulum whose restoring torque has two components: a parametrically modulated torque $\propto [i_{\rm dc}+i_{\rm ac}\cos(\omega t)]\sin x$
stemming from the gate-controlled critical current in leg 1, and  an additional static phase-shifted restoring torque $i_{c,t}\sin(x+\phi)$ from leg 2. If $\phi=0$ (or if branch 2 had $I_{s,2}\propto \sin x$), then the two torques combine into a single effective coefficient multiplying $\sin x$, and the distinct symmetry-breaking structure used in the main text would be lost.

\section{Reduction of a gated single-loop dc SQUID at finite applied flux to an Effective Parametrically Driven Pendulum Equation}
\label{sec:supp_singleloop_gate_flux}

In this section we extend the single-loop reduction of Sec.~\ref{sec:supp_singleloop_gate} to the case of a nonzero applied flux $\Phi_{\rm ext}\neq 0$. We show that, in the same regime assumed in the main text (negligible loop inductance and a common voltage across the two junctions), the dynamics still reduces to a single collective phase variable. The only modification is that the phase shift entering the static channel
$i_{c,t}\sin(x+\phi)$ is renormalized by the applied flux.

\subsection{Circuit model}

We consider a standard dc SQUID: two Josephson branches $j=1,2$ in parallel, forming a superconducting loop and biased by a total current $I_b$. The gauge-invariant phase drops across the branches are
$\varphi_1(t_{\rm phys})$ and $\varphi_2(t_{\rm phys})$.
Each branch is modeled by an RCSJ element with capacitance $C_j$ and shuntresistance $R_j$,
\begin{equation}
I_j
= C_j\frac{\Phi_0}{2\pi}\,\ddot\varphi_j
+ \frac{1}{R_j}\frac{\Phi_0}{2\pi}\,\dot\varphi_j
+ I_{s,j}(\varphi_j,t_{\rm phys}),
\qquad j=1,2,
\label{eq:RCSJ_each_flux}
\end{equation}
and current conservation gives
\begin{equation}
I_b = I_1+I_2.
\label{eq:Ib_split_flux}
\end{equation}
The Josephson voltage relations are
\begin{equation}
V_j(t_{\rm phys})=\frac{\Phi_0}{2\pi}\,\dot\varphi_j(t_{\rm phys}),
\qquad j=1,2.
\label{eq:JJ_voltage_flux}
\end{equation}

As in Sec.~\ref{sec:supp_singleloop_gate}, to obtain the effective equation
used in the main text we assume two distinct sinusoidal supercurrent channels:
\begin{align}
I_{s,1}(\varphi_1,t_{\rm phys})
&= I_{c,1}(t_{\rm phys})\,\sin\varphi_1,
\qquad
I_{c,1}(t_{\rm phys})
= I_{c,1}^{(0)} + \delta I_{c,1}\cos(\Omega t_{\rm phys}),
\label{eq:gate_mod_Ic_flux}
\\
I_{s,2}(\varphi_2)
&= I_{c,t}\,\sin(\varphi_2+\phi).
\label{eq:phi_shift_CPR_flux}
\end{align}
Here $\phi$ is a built-in phase offset of the second channel (e.g. a
$\varphi_0$-junction contribution or any microscopic mechanism producing $\sin(\varphi+\phi)$ at $\Phi=0$). Importantly, if leg 2 had the ordinary CPR $I_{s,2}\propto \sin\varphi_2$ (i.e. $\phi=0$ and no other phase-bias mechanism), then the two branches would not generate a distinct ``second channel'' in the single-phase reduction: they would simply renormalize the coefficient of $\sin x$. Thus, the appearance of a term $I_{c,t}\sin(x+\phi)$ in the final single-phase equation requires a genuinely phase-shifted contribution.

\subsection{Fluxoid quantization and the small-inductance limit}

The gauge-invariant phase difference around the loop is constrained by fluxoid
quantization,
\begin{equation}
\varphi_1-\varphi_2
= \frac{2\pi}{\Phi_0}\Big(\Phi_{\rm ext}+L I_{\rm circ}\Big),
\label{eq:fluxoid_general_flux}
\end{equation}
where $L$ is the loop inductance and $I_{\rm circ}$ is the circulating current.
In the regime assumed in the main paper and throughout this reduction,
\begin{equation}
L\to 0,
\qquad\text{so that}\qquad
\varphi_1-\varphi_2=\delta_\Phi,
\label{eq:fluxoid_L0_flux}
\end{equation}
with the flux-imposed phase bias
\begin{equation}
\delta_\Phi \equiv \frac{2\pi \Phi_{\rm ext}}{\Phi_0}.
\label{eq:deltaPhi_def}
\end{equation}
Equation~\eqref{eq:fluxoid_L0_flux} is algebraic: the applied flux fixes the difference $\varphi_1-\varphi_2$ at all times. Consequently, the two junction voltages are identical:
\begin{equation}
\dot\varphi_1-\dot\varphi_2=\frac{d}{dt_{\rm phys}}(\delta_\Phi)=0
\quad\Longrightarrow\quad
V_1(t_{\rm phys})=V_2(t_{\rm phys}),
\label{eq:common_voltage_flux}
\end{equation}
consistent with the parallel circuit topology.

\subsection{Single-phase reduction at finite flux}

To reduce the dynamics to a single variable, we choose the collective phase
to be one of the branch phases, for concreteness
\begin{equation}
x(t_{\rm phys}) \equiv \varphi_1(t_{\rm phys}).
\label{eq:x_def_flux}
\end{equation}
Then the second branch phase is fixed by the flux constraint
Eq.~\eqref{eq:fluxoid_L0_flux}:
\begin{equation}
\varphi_2(t_{\rm phys}) = \varphi_1(t_{\rm phys})-\delta_\Phi
= x(t_{\rm phys})-\delta_\Phi.
\label{eq:gamma2_in_terms_of_x_flux}
\end{equation}
Substituting Eq.~\eqref{eq:gamma2_in_terms_of_x_flux} into the supercurrent of
branch 2 gives
\begin{equation}
I_{s,2}(\varphi_2)
= I_{c,t}\sin(\varphi_2+\phi)
= I_{c,t}\sin\!\big(x+\phi-\delta_\Phi\big).
\label{eq:Is2_subst_flux}
\end{equation}
Thus, at finite flux the second channel retains the same functional form but
with an effective phase offset
\begin{equation}
\phi_{\rm eff} \equiv \phi-\delta_\Phi
= \phi - \frac{2\pi \Phi_{\rm ext}}{\Phi_0}.
\label{eq:phi_eff_def}
\end{equation}

\subsection{Physical equation of motion for the collective phase}

We now sum the two RCSJ equations~\eqref{eq:RCSJ_each_flux} over $j=1,2$ and use
Eqs.~\eqref{eq:Ib_split_flux}, \eqref{eq:x_def_flux},
\eqref{eq:gamma2_in_terms_of_x_flux}, and \eqref{eq:Is2_subst_flux}.
The capacitive and resistive terms combine into effective parallel parameters
\begin{equation}
C_\Sigma \equiv C_1+C_2,
\qquad
\frac{1}{R_\Sigma}\equiv \frac{1}{R_1}+\frac{1}{R_2}.
\label{eq:Reff_Ceff_flux}
\end{equation}
We obtain the physical-time equation for the single phase $x$:
\begin{equation}
C_\Sigma\frac{\Phi_0}{2\pi}\,\ddot x
+\frac{1}{R_\Sigma}\frac{\Phi_0}{2\pi}\,\dot x
+\Big[I_{c,1}^{(0)}+\delta I_{c,1}\cos(\Omega t_{\rm phys})\Big]\sin x
+ I_{c,t}\sin\!\big(x+\phi-\delta_\Phi\big)
= I_b,
\label{eq:phys_EOM_flux}
\end{equation}
where dots denote $d/dt_{\rm phys}$.

Eq. ~\eqref{eq:phys_EOM_flux} is the finite-flux counterpart of
Eq.~\eqref{eq:phys_EOM} in Sec.~\ref{sec:supp_singleloop_gate}. It differs only by the replacement $\phi\to\phi-\delta_\Phi$.

\subsection{Dimensionless units and mapping to the effective phase equation of motion}

We nondimensionalize Eq.~\eqref{eq:phys_EOM_flux} using the same conventions as
in Sec.~\ref{sec:supp_singleloop_gate}.
Choose a reference current scale $I_0$ (e.g. $I_0=I_{c,1}^{(0)}$) and define the
reference plasma frequency
\begin{equation}
\omega_{p0}\equiv
\sqrt{\frac{2\pi I_0}{\Phi_0\,C_\Sigma}}.
\label{eq:wp0_def_flux}
\end{equation}
Introduce dimensionless time $t=\omega_{p0}t_{\rm phys}$ and parameters
\begin{equation}
\beta \equiv \frac{1}{R_\Sigma C_\Sigma \omega_{p0}},
\qquad
\omega\equiv \frac{\Omega}{\omega_{p0}}.
\label{eq:beta_omega_def_flux}
\end{equation}
Define dimensionless current amplitudes
\begin{equation}
i_{\rm dc}\equiv \frac{I_{c,1}^{(0)}}{I_0},
\qquad
i_{\rm ac}\equiv \frac{\delta I_{c,1}}{I_0},
\qquad
i_{c,t}\equiv \frac{I_{c,t}}{I_0},
\label{eq:dimensionless_Ic_flux}
\end{equation}
and the bias convention
\begin{equation}
i_b \equiv \frac{2 I_b}{I_0}
\qquad \Longleftrightarrow \qquad
\frac{I_b}{I_0}=\frac12\,i_b.
\label{eq:ib_convention_flux}
\end{equation}
Dividing Eq.~\eqref{eq:phys_EOM_flux} by $I_0$ and rewriting in dimensionless
time yields
\begin{equation}
\ddot x + \beta \dot x
+ \big[i_{\mathrm{dc}}+i_{\mathrm{ac}}\cos(\omega t)\big]\sin x
+ i_{c,t} \sin\!\big(x+\phi-\delta_\Phi\big)
= \frac{1}{2} i_b ,
\label{eq:EOM_final_supp_flux}
\end{equation}
where dots now denote $d/dt$ with $t=\omega_{p0}t_{\rm phys}$.

Eq.~\eqref{eq:EOM_final_supp_flux} is therefore identical to
Eq.~\eqref{eq:EOM_supp} of the zero-flux case upon replacement
\begin{equation}
\phi \ \longrightarrow\ \phi_{\rm eff}\equiv \phi-\delta_\Phi
= \phi - \frac{2\pi \Phi_{\rm ext}}{\Phi_0}.
\label{eq:phi_to_phi_eff}
\end{equation}
In other words, a nonzero applied flux simply shifts the phase offset of the static channel.

\subsection{Remarks on the validity of the single-phase reduction}

The reduction to a single collective phase variable relies crucially on the small-inductance limit in Eq.~\eqref{eq:fluxoid_L0_flux}. If $L$ is not negligible, then Eq.~\eqref{eq:fluxoid_general_flux} contains the circulating current $I_{\rm circ}$ and the phase difference $\varphi_1-\varphi_2$ becomes a dynamical degree of freedom. In that case, one generally must work with
two coupled variables (for example, the sum and difference phases
$\varphi_\pm\equiv(\varphi_1\pm\varphi_2)/2$) rather than a single equation for $x$. Therefore, Eq.~\eqref{eq:EOM_final_supp_flux} should be viewed as controlled when the loop is sufficiently ``stiff'' (geometric flux bias dominates and the inductive back action is negligible).

\section{Reduction of the Double-Loop SQUID to an Effective
Parametrically Driven Pendulum Equation}
\label{sec:SM_reduction}

\subsection{Circuit model}
We consider the double-loop device shown in Fig.~3(b) of the main text.
Electrically, it is a two-terminal superconducting circuit connecting the same left and right superconducting electrodes by three parallel Josephson branches:
(i) an upper conventional Josephson junction (JJ), (ii) a middle ``anomalous'' (or ``topological'') JJ with an intrinsic phase shift $\phi$, and (iii) a lower conventional JJ. Two loops are formed between these branches and are threaded
by external magnetic fluxes $\Phi_1(t)$ and $\Phi_2(t)$ (see Fig.~3(b)).

The purpose of this section is to derive, from the full multi-branch RCSJ description and the fluxoid constraints, a single effective phase equation of motion of the form
\begin{equation}
\ddot x + \beta \dot x
+ \big[i_{\mathrm{dc}}+i_{\mathrm{ac}}\cos(\omega t)\big]\sin x
+ i_{c,t}\,\sin(x+\phi)
= \frac{1}{2}\,i_b,
\label{eq:EOM_SM_target}
\end{equation}
which is Eq.~(1) in the main text. Here and below, overdots denote derivatives with respect to dimensionless time, and all currents are normalized by a convenient current scale $I_0$.

\subsection{Gauge-invariant phase differences and Josephson relations}
\label{sec:SM_gaugephase}
Let $\varphi_j(t)$ denote the gauge-invariant phase difference across junction $j$ ($j=1,2,3$ corresponding to upper, middle, lower branches). It is defined as
\begin{equation}
\varphi_j(t) = \theta_L(t)-\theta_R(t)
-\frac{2\pi}{\Phi_0}\int_{\mathcal{C}_j}\mathbf{A}(\mathbf{r},t)\cdot d\mathbf{l},
\label{eq:SM_gaugeinv_phase}
\end{equation}
where $\theta_{L,R}$ are the superconducting phases of the left/right
electrodes, $\mathbf{A}$ is the vector potential, $\mathcal{C}_j$ is a contour
through the $j$th junction, and $\Phi_0=h/2e$ is the superconducting flux
quantum.

The voltage across junction $j$ is related to $\varphi_j$ by the second Josephson
relation,
\begin{equation}
V_j(t)=\frac{\Phi_0}{2\pi}\,\frac{d\varphi_j}{dt_{\rm phys}},
\label{eq:SM_josephson_voltage}
\end{equation}
where $t_{\rm phys}$ is physical time. Since all three junctions connect the same two electrodes in parallel, they share the same two-terminal voltage
$V(t)$, hence
\begin{equation}
V_1(t)=V_2(t)=V_3(t)\equiv V(t)
\quad \Longrightarrow \quad
\dot\varphi_1=\dot\varphi_2=\dot\varphi_3
\ \ \text{(in physical time)}.
\label{eq:SM_commonV}
\end{equation}
Therefore, differences $\varphi_1-\varphi_2$ and $\varphi_2-\varphi_3$ are not
independent dynamical variables: they are (to leading order) fixed by loop
fluxoid constraints (up to $2\pi$ integers), as shown next.

\subsection{Two-loop fluxoid quantization constraints}
\label{sec:SM_fluxoid}
We assume the standard ``small-inductance'' limit for each loop:
the geometric inductances are sufficiently small that self-induced flux is
negligible compared to the applied flux, so that the loop constraints are set
by the externally applied fluxes $\Phi_1(t)$ and $\Phi_2(t)$ alone.

For each loop, fluxoid quantization gives a constraint for the sum of
gauge-invariant phase drops around the loop. For the upper loop (between
junctions 1 and 2, threaded by $\Phi_2$) and lower loop (between junctions 2 and
3, threaded by $\Phi_1$), one can write
\begin{align}
\varphi_1(t)-\varphi_2(t)
&= \frac{2\pi}{\Phi_0}\,\Phi_2(t) + 2\pi n_2,
\label{eq:SM_constraint_upper}
\\
\varphi_2(t)-\varphi_3(t)
&= \frac{2\pi}{\Phi_0}\,\Phi_1(t) + 2\pi n_1,
\label{eq:SM_constraint_lower}
\end{align}
where $n_{1,2}\in\mathbb{Z}$ label fluxoid branches. In what follows we absorb
$2\pi n_{1,2}$ into the definition of the phases and work on a fixed branch,
so that
\begin{equation}
\varphi_1-\varphi_2 = f_2(t),
\qquad
\varphi_2-\varphi_3 = f_1(t),
\qquad
f_j(t)\equiv\frac{2\pi}{\Phi_0}\,\Phi_j(t)\ \ (j=1,2).
\label{eq:SM_define_f}
\end{equation}

A convenient choice is to take the middle-junction phase as the collective
coordinate,
\begin{equation}
x(t)\equiv \varphi_2(t).
\label{eq:SM_define_x}
\end{equation}
Then Eqs.~\eqref{eq:SM_define_f} imply
\begin{equation}
\varphi_1(t)=x(t)+f_2(t),
\qquad
\varphi_3(t)=x(t)-f_1(t).
\label{eq:SM_gamma123}
\end{equation}
Thus the full three-junction problem reduces to a single dynamical variable
$x(t)$, with the fluxes entering as externally controlled offsets.

\subsection{RCSJ equations for the three branches and current conservation}
\label{sec:SM_RCSJ}
We model each branch by a resistively and capacitively shunted junction (RCSJ).
The total current through branch $j$ is
\begin{equation}
I_j
=
I_{c,j}\, \sin\big(\varphi_j+\delta_j\big)
+\frac{V}{R_j}
+C_j\,\frac{dV}{dt_{\rm phys}},
\label{eq:SM_RCSJ_general}
\end{equation}
where $I_{c,j}$ is the critical current, $R_j$ and $C_j$ are shunt resistance
and capacitance, and $\delta_j$ is a possible intrinsic phase shift.
For the two conventional junctions we set $\delta_1=\delta_3=0$, while for the
anomalous (middle) junction we set $\delta_2=\phi$:
\begin{equation}
\delta_1=\delta_3=0,
\qquad
\delta_2=\phi.
\label{eq:SM_delta}
\end{equation}
Using the common voltage relation \eqref{eq:SM_josephson_voltage}--\eqref{eq:SM_commonV},
\begin{equation}
V(t)=\frac{\Phi_0}{2\pi}\,\dot x(t)\quad\text{(physical time)},
\qquad
\frac{dV}{dt_{\rm phys}}=\frac{\Phi_0}{2\pi}\,\ddot x(t).
\label{eq:SM_V_in_terms_of_x}
\end{equation}

The external bias current $I_b$ injected into the two-terminal device is
distributed among the three branches,
\begin{equation}
I_b(t)=I_1(t)+I_2(t)+I_3(t).
\label{eq:SM_current_conservation}
\end{equation}
Substituting \eqref{eq:SM_gamma123}--\eqref{eq:SM_V_in_terms_of_x} into the
branch currents \eqref{eq:SM_RCSJ_general} yields a closed equation for $x(t)$.

\subsection{Symmetric outer junctions and effective ``parametric'' restoring term}
\label{sec:SM_parametric}
To obtain a compact effective equation, we assume that the upper and lower junctions are identical:
\begin{equation}
I_{c,1}=I_{c,3}\equiv I_{c0},
\qquad
R_1=R_3\equiv R_0,
\qquad
C_1=C_3\equiv C_0.
\label{eq:SM_identical_outer}
\end{equation}
We keep the middle junction parameters $(I_{c,t},R_t,C_t)$ distinct.

Using \eqref{eq:SM_gamma123}, the superconducting currents are
\begin{align}
I_1^{(J)} &= I_{c0}\sin\big(x+f_2(t)\big),
\\
I_2^{(J)} &= I_{c,t}\sin\big(x+\phi\big),
\\
I_3^{(J)} &= I_{c0}\sin\big(x-f_1(t)\big).
\end{align}
Thus, the pair of conventional branches contributes
\begin{equation}
I_1^{(J)}+I_3^{(J)}
=
I_{c0}\sin(x+f_2)+I_{c0}\sin(x-f_1).
\label{eq:SM_pair_general}
\end{equation}

A particularly important (and experimentally natural) operation mode is the symmetric flux drive in which the two loops are driven equally,
\begin{equation}
f_1(t)=f_2(t)\equiv f(t)
\quad \Longleftrightarrow \quad
\Phi_1(t)=\Phi_2(t).
\label{eq:SM_symmetric_drive}
\end{equation}
Then Eq.~\eqref{eq:SM_pair_general} simplifies by the trigonometric identity
$\sin(x+f)+\sin(x-f)=2\sin x\cos f$:
\begin{equation}
I_1^{(J)}+I_3^{(J)}
=
2I_{c0}\cos\big(f(t)\big)\,\sin x.
\label{eq:SM_pair_parametric_exact}
\end{equation}
Equation \eqref{eq:SM_pair_parametric_exact} explicitly shows that the
conventional pair acts as a restoring torque $\propto \sin x$ whose strength is controlled by the flux through the multiplicative factor $2I_{c0}\cos f(t)$, i.e.\ a parametric modulation of the effective ``gravity'' term in the pendulum analogy.

We now implement a dc+ac flux protocol
\begin{equation}
f(t)=f_{\rm dc}+f_{\rm ac}\cos(\Omega t_{\rm phys}),
\label{eq:SM_flux_protocol}
\end{equation}
where $f_{\rm dc}=(2\pi/\Phi_0)\Phi_{\rm dc}$, and $f_{\rm ac}=(2\pi/\Phi_0)\Phi_{\rm ac}$
is the ac amplitude. In the weak-modulation regime $|f_{\rm ac}|\ll 1$, we expand
\begin{equation}
\cos f(t)
=
\cos f_{\rm dc}
-
f_{\rm ac}\sin f_{\rm dc}\cos(\Omega t_{\rm phys})
+O(f_{\rm ac}^2).
\label{eq:SM_cos_expand}
\end{equation}
Substituting \eqref{eq:SM_cos_expand} into \eqref{eq:SM_pair_parametric_exact} gives
\begin{equation}
I_1^{(J)}+I_3^{(J)}
\approx
\underbrace{2I_{c0}\cos f_{\rm dc}}_{\displaystyle I_{\rm dc}}\sin x
+
\underbrace{\big(-2I_{c0}f_{\rm ac}\sin f_{\rm dc}\big)}_{\displaystyle I_{\rm ac}}
\cos(\Omega t_{\rm phys})\,\sin x
\ +\ O(f_{\rm ac}^2).
\label{eq:SM_pair_parametric_expand}
\end{equation}
Hence the conventional pair indeed generates the structure
$\big[I_{\rm dc}+I_{\rm ac}\cos(\Omega t_{\rm phys})\big]\sin x$.

The middle junction contributes the additional term
\begin{equation}
I_2^{(J)}=I_{c,t}\sin(x+\phi),
\label{eq:SM_anomalous_term}
\end{equation}
which can be viewed mechanically as a second restoring torque (or ``spring'')
whose equilibrium is shifted by $\phi$.

\subsection{Effective inertial and damping terms}
\label{sec:SM_damping_inertia}
The dissipative (resistive) currents add because the branches are in parallel:
\begin{equation}
I^{(R)}= \frac{V}{R_0}+\frac{V}{R_t}+\frac{V}{R_0}
=
G_\Sigma\,V,
\qquad
G_\Sigma\equiv \frac{2}{R_0}+\frac{1}{R_t}.
\label{eq:SM_Gsum}
\end{equation}
Similarly, capacitive currents add:
\begin{equation}
I^{(C)}=
C_0\frac{dV}{dt_{\rm phys}}+C_t\frac{dV}{dt_{\rm phys}}+C_0\frac{dV}{dt_{\rm phys}}
=
C_\Sigma\,\frac{dV}{dt_{\rm phys}},
\qquad
C_\Sigma\equiv 2C_0+C_t.
\label{eq:SM_Csum}
\end{equation}
Using \eqref{eq:SM_V_in_terms_of_x}, these become
\begin{equation}
I^{(R)}=\frac{\Phi_0}{2\pi}G_\Sigma\,\dot x,
\qquad
I^{(C)}=\frac{\Phi_0}{2\pi}C_\Sigma\,\ddot x
\quad \text{(physical time)}.
\label{eq:SM_RC_in_x}
\end{equation}

\subsection{Closed physical-time equation for $x(t)$}
\label{sec:SM_physical_equation}
Collecting all contributions, the current balance \eqref{eq:SM_current_conservation}
yields (to leading order in $f_{\rm ac}$)
\begin{align}
\frac{\Phi_0}{2\pi}C_\Sigma\,\ddot x
+\frac{\Phi_0}{2\pi}G_\Sigma\,\dot x
+\Big[I_{\rm dc}+I_{\rm ac}\cos(\Omega t_{\rm phys})\Big]\sin x
+I_{c,t}\sin(x+\phi)
= I_b,
\label{eq:SM_physical_x_equation}
\end{align}
where $I_{\rm dc}$ and $I_{\rm ac}$ are defined in \eqref{eq:SM_pair_parametric_expand}.
Equation \eqref{eq:SM_physical_x_equation} is the desired single-coordinate
description in physical units.

\subsection{Dimensionless units and mapping to the effective phase equation of motion}
\label{sec:SM_nondim}
We now introduce a dimensionless time and currents to obtain the compact form used in the main text.

A natural current scale for the symmetric device is the sum of the two outer critical currents
\begin{equation}
I_0 \equiv 2I_{c0}.
\label{eq:SM_I0_def}
\end{equation}
We define a reference plasma frequency associated with the total capacitance $C_\Sigma$ and current scale $I_0$,
\begin{equation}
\omega_{p0}\equiv \sqrt{\frac{2\pi I_0}{\Phi_0 C_\Sigma}}.
\label{eq:SM_omega_p0}
\end{equation}
Introduce dimensionless time
\begin{equation}
t \equiv \omega_{p0}\,t_{\rm phys},
\label{eq:SM_dimless_time}
\end{equation}
so that $d/dt_{\rm phys}=\omega_{p0}\,d/dt$ and $d^2/dt_{\rm phys}^2=\omega_{p0}^2\,d^2/dt^2$.

Define the dimensionless damping constant
\begin{equation}
\beta \equiv \frac{G_\Sigma}{C_\Sigma\,\omega_{p0}},
\label{eq:SM_beta_def}
\end{equation}
and the dimensionless drive frequency
\begin{equation}
\omega \equiv \frac{\Omega}{\omega_{p0}}.
\label{eq:SM_omega_def}
\end{equation}

We normalize the effective dc and ac amplitudes by $I_0$:
\begin{equation}
i_{\rm dc}\equiv \frac{I_{\rm dc}}{I_0},
\qquad
i_{\rm ac}\equiv \frac{I_{\rm ac}}{I_0},
\qquad
i_{c,t}\equiv \frac{I_{c,t}}{I_0}.
\label{eq:SM_idc_iac_ict_def}
\end{equation}
Finally, the normalized bias current entering Eq.~\eqref{eq:SM_physical_x_equation} is
\begin{equation}
i_b^{(0)}\equiv \frac{I_b}{I_0}.
\label{eq:SM_ib0_def}
\end{equation}

\paragraph*{Dimensionless equation.}
Dividing Eq.~\eqref{eq:SM_physical_x_equation} by $I_0$ and using the above
definitions, we obtain
\begin{equation}
\ddot x + \beta \dot x
+ \big[i_{\rm dc}+i_{\rm ac}\cos(\omega t)\big]\sin x
+ i_{c,t}\sin(x+\phi)
= i_b^{(0)}.
\label{eq:SM_dimless_equation}
\end{equation}

In some conventions, one defines the plotted bias current as
$i_b \equiv I_b/I_{c0}$, i.e.\ normalized by a single outer junction
critical current. Since $I_0=2I_{c0}$, we then have
\begin{equation}
i_b^{(0)}=\frac{I_b}{I_0}=\frac{I_b}{2I_{c0}}=\frac{1}{2}\,\frac{I_b}{I_{c0}}
=\frac{1}{2}\,i_b.
\label{eq:SM_half_factor}
\end{equation}
Substituting \eqref{eq:SM_half_factor} into \eqref{eq:SM_dimless_equation}
reproduces exactly Eq.~\eqref{eq:EOM_SM_target} used in the main text.

\subsection{Remarks on validity and generalizations}
\label{sec:SM_validity}

The reduction \eqref{eq:SM_constraint_upper}--\eqref{eq:SM_gamma123} assumes negligible loop self-inductances. If inductances are not negligible, then the fluxoid constraints acquire additional terms proportional to circulating currents, and $f_1,f_2$ become coupled to the dynamics. In that case one must solve a larger set of equations (phases plus loop currents). The present limit is appropriate for sufficiently small loops or sufficiently weak circulating
currents.

The clean parametric form \eqref{eq:SM_pair_parametric_exact} arises for $f_1(t)=f_2(t)$. If $f_1\neq f_2$, the conventional pair yields
$\sin(x+f_2)+\sin(x-f_1)$, which can be rewritten as a combination of $\sin x$ and $\cos x$ with time-dependent coefficients. This leads to a more general effective CPR containing both first-harmonic sine and cosine terms. The symmetric protocol is the simplest route to the compact form used in the main text.

The identification of $i_{\rm dc}$ and $i_{\rm ac}$ via \eqref{eq:SM_pair_parametric_expand} assumes $|f_{\rm ac}|\ll 1$. For larger $f_{\rm ac}$, higher harmonics in $\cos f(t)$ appear and generate additional time-dependent terms (including higher Fourier components), which can be kept systematically if needed.

Eq. \eqref{eq:EOM_SM_target} is mechanically equivalent to a damped pendulum with a gravitational torque whose strength is modulated in time, $\propto [i_{\rm dc}+i_{\rm ac}\cos(\omega t)]\sin x$, and an additional phase-shifted restoring torque $i_{c,t}\sin(x+\phi)$, which acts like a second torsional spring with equilibrium rotated by $\phi$.

\section{Kapitza averaging with AC current: derivation of the effective CPR and frequency-dependent rectification amplitude $\eta(\omega)$}

We start from the driven equation of motion
\begin{equation}
\ddot x + \beta \dot x
+ \big[i_{\mathrm{dc}}+i_{\mathrm{ac}}\cos(\omega t)\big]\sin x
+ i_{c,t}\sin(x+\phi)
= \frac{1}{2}i_b .
\label{eq:EOM}
\end{equation}
Following the Kapitza ideology, we decompose
\begin{equation}
x(t)=X(t)+\xi(t),
\end{equation}
where $X(t)$ is slow and $\xi(t)$ is a small fast correction, $|\xi|\ll 1$.
Expanding Eq.~\eqref{eq:EOM} up to $O(\xi^2)$ gives
\begin{align}
\ddot X+\ddot \xi + \beta(\dot X+\dot \xi)
&+ \big[i_{\mathrm{dc}}+i_{\mathrm{ac}}\cos(\omega t)\big]
\Big(\sin X + \xi \cos X -\tfrac{1}{2}\xi^2\sin X\Big)
\nonumber\\
&+ i_{c,t}\Big(\sin(X+\phi)+\xi\cos(X+\phi)-\tfrac{1}{2}\xi^2\sin(X+\phi)\Big)
=\frac{1}{2}i_b.
\label{eq:expanded2}
\end{align}

On the fast time scale, we neglect $\ddot X$ and $\dot X$ and retain the terms
linear in $\xi$ and proportional to the oscillatory part
$\delta c(t)\equiv i_{\mathrm{ac}}\cos(\omega t)$. This yields
\begin{equation}
\ddot \xi + \beta \dot \xi + \omega_0^2(X)\,\xi
\simeq
-i_{\mathrm{ac}}\cos(\omega t)\,\sin X,
\label{eq:xi-eq-corr}
\end{equation}
where
\begin{equation}
\omega_0^2(X)\equiv i_{\mathrm{dc}}\cos X + i_{c,t}\cos(X+\phi).
\label{eq:omega0}
\end{equation}
The steady periodic solution of Eq.~\eqref{eq:xi-eq-corr} can be written as
\begin{equation}
\xi(t)=\Re\!\left[\tilde\xi\,e^{i\omega t}\right],
\qquad
\tilde\xi=
-\frac{i_{\mathrm{ac}}\sin X}{\omega_0^2(X)-\omega^2+i\beta\omega}.
\label{eq:xi-sol-corr}
\end{equation}
Hence, the relevant time averages are
\begin{align}
\langle \xi\rangle_t &= 0,
\qquad
\langle \xi^2\rangle_t = \frac{|\tilde\xi|^2}{2}
=
\frac{i_{\mathrm{ac}}^2\sin^2 X}{2D(X,\omega)},
\label{eq:xi2-corr}
\\
\Big\langle \cos(\omega t)\,\xi(t)\Big\rangle_t
&=\frac{1}{2}\Re(\tilde\xi)
=
-\frac{i_{\mathrm{ac}}\sin X}{2}\,
\frac{\omega_0^2(X)-\omega^2}{D(X,\omega)},
\label{eq:cosxi-corr}
\end{align}
with the standard Lorentzian denominator
\begin{equation}
D(X,\omega)\equiv \big(\omega_0^2(X)-\omega^2\big)^2+(\beta\omega)^2.
\label{eq:Ddef}
\end{equation}

Averaging Eq.~\eqref{eq:expanded2} over the fast oscillations and keeping all
terms up to $O(i_{\mathrm{ac}}^2)$, we obtain
\begin{align}
\ddot X + \beta \dot X
&+ i_{\mathrm{dc}}\sin X + i_{c,t}\sin(X+\phi)
\nonumber\\
&\underbrace{+\ i_{\mathrm{ac}}\cos X\Big\langle \cos(\omega t)\,\xi(t)\Big\rangle_t}_{\text{cross term }O(i_{\mathrm{ac}}^2)}
\underbrace{-\ \frac{1}{2}\langle\xi^2\rangle_t\Big[i_{\mathrm{dc}}\sin X+i_c\sin(X+\phi)\Big]}_{O(i_{\mathrm{ac}}^2)}
=\frac{1}{2}I_b.
\label{eq:slow-corr}
\end{align}
The second line contains two distinct $O(\Phi_{\mathrm{ac}}^2)$ contributions. Importantly, the cross term survives at the unmodulated equilibrium and must be retained for a consistent Kapitza reduction.

For the diode effect we focus on the vicinity of the static equilibrium
$x_0$ at zero bias (or small $I_b$), defined by the unmodulated CPR
\begin{equation}
i_{\mathrm{dc}}\sin x_0+i_c\sin(x_0+\phi)=0,
\qquad
x_0=\operatorname{atan2}\!\big(-i_{c,t}\sin\phi,\ i_{\mathrm{dc}}+i_{c,t}\cos\phi\big).
\label{eq:x0}
\end{equation}
Near equilibrium $x_0$, the contribution is proportional to
$\langle\xi^2\rangle_t[\Phi_{\mathrm{dc}}\sin X+i_c\sin(X+\phi)]$
is proportional to the unmodulated force and thus vanishes at $X=x_0$.
The leading symmetry-breaking $O(\Phi_{\mathrm{ac}}^2)$ correction therefore originates from the cross term. Using Eq.~\eqref{eq:cosxi-corr} and $\sin X\cos X=\tfrac12\sin2X$,
we can encode the averaged dynamics in an effective static CPR
\begin{equation}
I_s^{\mathrm{eff}}(X)=A_1\sin X+B_1\cos X + A_2(\omega)\,\sin 2X,
\label{eq:CPR-eff}
\end{equation}
with
\begin{equation}
A_1=i_{\mathrm{dc}}+i_{c,t}\cos\phi,\qquad B_1=i_{c,t}\sin\phi,
\label{eq:A1B1}
\end{equation}
Approximating the slowly varying prefactor
$[\omega_0^2(X)-\omega^2]/D(X,\omega)$ by its value at $X=x_0$, we obtain
\begin{equation}
A_2(\omega)=
-\frac{i_{\mathrm{ac}}^2}{4}\,
\frac{\omega_0^2(x_0)-\omega^2}{\big(\omega_0^2(x_0)-\omega^2\big)^2+(\beta\omega)^2}.
\label{eq:A2-corr}
\end{equation}
Introducing
\begin{equation}
Q=\sqrt{A_1^2+B_1^2},\qquad \delta = \arctan\!\frac{B_1}{A_1},
\label{eq:Rdelta}
\end{equation}
the first harmonic becomes $Q\sin(X+\delta)$.

For $|A_2|\ll R$, the extrema of
$I_s^{\mathrm{eff}}(X)=Q\sin(X+\delta)+A_2\sin2X$ give
\begin{equation}
I_c^+ \approx Q + A_2(\omega)\,\sin(2\delta),
\qquad
I_c^- \approx -Q + A_2(\omega)\,\sin(2\delta),
\label{eq:Icpm-corr}
\end{equation}
and therefore the diode efficiency;
\begin{equation}
\eta(\omega)=\frac{I_c^+-|I_c^-|}{I_c^+ + |I_c^-|}
\approx
\frac{A_2(\omega)}{Q}\,\sin(2\delta).
\label{eq:eta-gen-corr}
\end{equation}
Substituting Eq.~\eqref{eq:A2-corr} finally yields the corrected analytic
frequency dependence
\begin{equation}
\eta(\omega)
\;\approx\;
-\frac{i_{\mathrm{ac}}^2}{4R}\,
\frac{\omega_0^2(x_0)-\omega^2}
     {\big[\omega_0^2(x_0)-\omega^2\big]^2+(\beta\omega)^2}
\,\sin 2\delta
\label{eq:eta-final-correct}
\end{equation}
with
\begin{equation}
\omega_0^2(x_0)=i_{\mathrm{dc}}\cos x_0 + i_{c,t}\cos(x_0+\phi),
\end{equation}
and $x_0,Q,\delta$ given in Eqs.~\eqref{eq:x0} and \eqref{eq:Rdelta}.

\end{document}